# Gibbs-Shannon Entropy and Related Measures: Tsallis Entropy

Garimella Rama Murthy,
Associate Professor,
IIIT---Hyderabad,
Gachibowli, HYDERABAD-32, AP, INDIA

**ABSTRACT**

In this research paper, it is proved that an approximation to Gibbs-Shannon entropy measure naturally leads to Tsallis entropy for the real parameter $q=2$. Several interesting measures based on the input as well as output of a discrete memoryless channel are provided and some of the properties of those measures are discussed. It is expected that these results will be of utility in Information Theoretic research.

## 1. Introduction:

From the considerations of statistical physics, J. Willard Gibbs proposed an interesting entropy measure. Independently, motivated by the desire to capture the uncertainity associated with a random variable, C.E. Shannon proposed an entropy measure. It is later realized that Gibbs and Shannon entropy measures are very closely related. From the considerations of statistical communication theory, defining mutual information ( based on the definition of conditional entropy measure ), Shannon successfully proved the channel coding Theorem ( which defined the limits of reliable communication from a source to destination over a noisy channel )[Ash]. Thus from the point of view of information theory, Shannon entropy measure became very important and useful.

Also, in recent years, Tsallis proposed an entropy measure generalizing the Boltzmann-Gibbs entropy measure. The author became interested in the relationship between Tsallis entropy and Shannon entropy. Under some conditions, the author proved that Shannon entropy leads to Tsallis entropy. As a natural generalization, the author proposed interesting measures defined on probability mass functions and the channel matrix of a Discrete Memoryless Channel (DMC).

This research paper is organized as follows. In Section 2, an approximation to Shannon entropy is discussed and the relationship to Tsallis entropy is proved. In Section 3, interesting measures on probability mass functions are proposed. Finally, conclusions are reported in Section 4.

## 2. Approximation to Gibbs-Shannon Entropy Measure : Tsallis Entropy:

It is well known that the Gibbs-Shannon entropy measure associated with a discrete random variable ( specified by the probability mass function $\{ \{p_i\} \, for \, 1 \leq i \leq M \}$ ) is given by

$$H(X) = - \sum_{i=1}^{M} p_i \, log_2 \, p_i \text{ bits. } \ldots\ldots\ldots\ldots..(2.1)$$

Also, in recent years, Tsallis proposed another entropy measure which in the case of a discrete random variable is given by

$$S_q(p) = \frac{1}{q-1} ( 1 - \sum_x ( p(x) )^q , \quad \ldots\ldots\ldots\ldots\ldots\ldots.(2.2)$$

where S denotes the entropy, p(.) is the probability mass function of interest and 'q' is a real parameter. In the limit as $q \to 1$, the normal Boltzmann—Gibbs entropy is recovered. The parameter 'q' is a measure of the non-extensivity of the system of interest.

The author became interested in knowing whether there is any relationship between Gibbs-Shannon entropy measure and Tsallis entropy under some conditions. This question naturally led to a discovery which is summarized in the following Lemma.

**Lemma 1**: Consider a discrete random variable X with finite support for the probability mass function. Under reasonable assumptions, we have that

$$H(X) \approx 1 - \sum_{i=1}^{M} p_i^2 = S_2(p) \quad \ldots\ldots\ldots\ldots(2.3)$$

**Proof**: From the basic theory of infinite series, we have the following:

For $|x| < 1$, we have the following

$$\log(1-x) = -x + \frac{x^2}{2} - \frac{x^3}{3} + \cdots + (-1)^{n-1}\frac{(-x)^n}{n} + \ldots$$

Let $p_i = (1 - q_i)$ with $0 < p_i < 1$. Then we have $0 < q_i < 1$. Thus, we have

$$\log(1 - q_i) = -q_i + \frac{q_i^2}{2} - \frac{q_i^3}{3} + \ldots \quad \ldots\ldots\ldots\ldots(2.4)$$

Now let us consider the entropy, H(X) of a discrete random variable, X which assumes finitely many values. We have that

$$H(X) = -\sum_{i=1}^{M} p_i \log_2 p_i \text{ bits } = -\sum_{i=1}^{M}(1 - q_i)\log_2(1 - q_i) \text{ bits}\ldots\ldots\ldots\ldots(2.5)$$

Now using the above infinite series and neglecting the terms $\frac{q_i^2}{2}, \frac{q_i^3}{3}$, and so on, We have that

$$H(X) \approx -\sum_{i=1}^{M}(1 - q_i)(-q_i) = \sum_{i=1}^{M} p_i(1 - p_i) = 1 - \sum_{i=1}^{M} p_i^2 \quad \text{Q.E.D.}$$

**Remark:** Thus, the square of the $L^2 - norm$ of the vector corresponding to the probability mass function ( of a discrete random variable ) is utilized to approximate the entropy of the discrete random variable. In summary, we have that

$$H(X) \approx f(p_1, p_2, \ldots, p_M) = 1 - \sum_{i=1}^{M} p_i^2 \quad \ldots\ldots\ldots\ldots(2.6)$$

Thus, an approximation to Gibbs-Shannon entropy naturally lead to the Tsallis entropy for the real parameter q=2. For continuous case i.e for probability density functions associated with continuous random variables, similar result can easily be derived and is avoided for brevity.

**Note:** If the logarithm is taken to a different base, a scaling constant should be included.

We would like to study the properties satisfied by the function f(.,.,,....) approximating the entropy. The following claim can easily be proved.

**Claim:** The maximum value of $f(p_1, p_2, \ldots, p_M)$ is attained when $\{p_i\}_{i=1}^{M}$

are all equal i.e. $p_i = \frac{1}{M}$ for $1 \leq i \leq M$.

Proof: Follows by the application of Lagrange Multipliers method. Detailed proof is avoided for brevity  Q.E.D

Thus, the maximum value of approximation to entropy of a discrete random variable assuming "M" values is $1 - \frac{1}{M}$.

- It is easy to see that this approximation to Gibbs-Shannon entropy satisfies only two (out of four) axioms satisfied by the Shannon entropy functional.

- **Remark:** As in the proof of above Lemma, it is possible to provide higher order approximations to Gibbs-Shannon entropy measure.

3. **Novel Measures on Probability Distributions:**

Shannon's entropy of a discrete random variable constitutes an important scalar valued mesure defined on the class of probability mass functions (of the discrete random variables). In contrast to the moments of discrete random variables, the entropy does not depend on the actual values assumed by the discrete random variable. Thus, one is naturally led to the definition of other measures associated with discrete random variable which depend only on the probability mass function (and not the values assumed by it).

- We first treat the probability mass function of a discrete random variable as a vector of probabilities. It should be kept in mind the M-dimensional probability vector (corresponding to 'M' values assumed by the discrete random variable) lies on a hyperplane in the "positive orthant" (of the M-dimensional Euclidean space) only. Also, as a natural generalization, we can also conceptualize "infinite dimensional" probability vector corresponding to a discrete random variable which assumes infinitely many values.

- Consider a "probability vector" (corresponding to the associated probability mass function..finite or infinite dimensional) and define the $L^p - norm$ of the vector (in the same manner as done in pure mathematics). Let

$$M_q = [\sum_{j=1}^{\infty}[p_X(j)]^q]^{\frac{1}{q}} \text{ for } q \geq 1 \ldots\ldots (2.7)$$

As discussed in [Rama], some interesting properties are satisfied by $M_q$.

It is elementary to see that such a measure can easily be related to Tsallis entropy. Specifically, we have that

$$M_q^q = 1 - (q-1)S_q(p). \quad \ldots\ldots\ldots\ldots(2.8)$$

- Based on the properties of $M_q$, it is easy to see that the probability mass function based infinite dimensional probability vectors always belong to discrete Hilbert space.

- Let us first consider the case where the support of the probability mass function is finite. The $L^2 - norm$ of the associated probability vector is

$$M_2 = [\sum_{j=1}^{M}[p_X(j)]^2]^{\frac{1}{2}} \ldots\ldots\ldots\ldots\ldots(2.9)$$

We reasoned in the previous section that such a measure naturally arises in approximating the Gibbs-Shannon entropy functional/measure (to a good degree of accuracy).

- Using similar approach, the conditional entropy can be approximated. Also, using the approximation for H(X|Y) / H(Y|X), H(X)/H(Y), the mutual information between the input and output of a Discrete Memoryless Channel (DMC) can be approximated. Details are avoided for brevity.

- Clearly, the expression in (2.3) is an interesting measure defined over the Vector, $\overline{p_X}$ representing the probability mass function. Thus one is naturally led to the definition of a quadratic form defined over the vector $\overline{p_X}$. Specifically, let us define quadratic forms associated with the channel matrix, Q ( of a DMC ).

$$\text{i.e. } \overline{p_X}^T \, Q \, \overline{p_X}.$$

Since $\overline{p_X}^T \, Q = \overline{p_Y}$, we readily have that

$$\overline{p_X}^T \, Q \, \overline{p_X} = \overline{p_Y}^T \, \overline{p_X} = \langle \overline{p_Y}, \overline{p_X} \rangle. \quad \ldots(2.10)$$

**Claim:** Thus the quadratic form associated with the channel matrix of a DMC represents the inner producy between the probability vectors $\overline{p_Y} \text{ and } \overline{p_X}$.

- It readily follows that in the case of "noiseless channel", we have that Q = I and thus the quadratic form becomes the "square of the Euclidean length ( $L^2$-norm ) of the probability vector. It is thus always positive.

- Hence we would like to study the properties of the quadratic form using the Inner product between two probability vectors ( namely the input and output probability vectors of a DMC ). In that effort, we would like to address the following question:

Q: How does the inner product of two probability vectors summarize the "similarity / dissimilarity" of probability mass functions?

***In this effort, we invoke the Cauchy-Schwartz inequality associated with bounding the inner-product between two vectors:

$$[\, \overline{p_X}^T \, \overline{p_Y} \,]^2 \leq [\, \sum_{i=1}^{M} p_X^2(i) \,] [\, \sum_{i=1}^{M} p_Y^2(i)] \ldots\ldots\ldots\ldots..(2.11)$$

- It is easily seen that the following holds true:

$$\sum_{i=1}^{M} p_X^2(i) = \begin{cases} 1 \text{ if } \overline{p_X} \text{ is degenerate} \\ < 1 \text{ if } \overline{p_X} \text{ is non} - \text{degenerate} \end{cases} \ldots\ldots..(2.12)$$

- Furthermore the minimum possible value of $\sum_{i=1}^{M} p_X^2(i)$ ( i.e. value of $\frac{1}{M}$ ) occurs when $p_X(i) = \frac{1}{M}$ for all $1 \leq i \leq M$.
- Also, it should be noted that the inequality in (2.11) reduces to equality only when

$$p_X(i) = p_Y(i) \; for \; 1 \leq i \leq M.$$

i.e the inner product between probability vectors $\overline{p_X}$ and $\overline{p_Y}$ attains the "maximum" value when they are both same ( equal).

- Suppose $\overline{p_X}$ is the invariant probability distribution ( also called as the steady state probability distribution ) of the homogeneous Discrete Time Markov Chain ( DTMC ) associated with the Channel matrix Q ( a stochastic matrix ). In this case, we have that
$$\overline{p_X^T} Q = \overline{p_X^T}. \qquad \ldots\ldots\ldots\ldots..(2.13)$$
Then the quadratic form associated with $\overline{p_X}$ becomes
$$\overline{p_X^T} \, Q \, \overline{p_X} = \overline{p_X^T} \, \overline{p_X} > 0.$$
Thus, the quadratic form attains the maximum value. Equivalently, we have that in this case, the value of the quadratic form is same as that in the case of a noiseless channel.

- In the same spirit of the definition of mutual information, let us define the following scalar "measure" between the input and output of a Discrete Memoryless Channel (DMC).
$$J(X;Y) = \overline{p_X^T} \, Q \, \overline{p_Y} = \overline{p_Y^T} \, \overline{p_Y}, \qquad \ldots\ldots..(2.14)$$
where $\overline{p_X}$ corresponds to the input probability vector (i.e the input probability mass function ) and $\overline{p_Y}$ corresponds to the output probability vector. Let us investigate some of the properties of the scalar measure $J(X;Y)$.

- (i) Since $\overline{p_X^T} Q = \overline{p_Y^T}$, we have that $\overline{p_Y^T} \, \overline{p_Y} > 0$.
  Also, we have that
  $$J(X;X) = = \overline{p_X^T} \, Q \, \overline{p_X} = \overline{p_Y^T} \, \overline{p_X} \geq 0.$$
  i.e. $J(X;X)$ is zero when the probability vectors $\overline{p_X}, \overline{p_Y}$ are orthogonal vectors ( as in the case of vector spaces ).

- (ii) $J(Y;X) = \overline{p_Y^T} \, Q \, \overline{p_X}$. Now substituting $\overline{p_Y^T} = \overline{p_X^T} Q$, we have that
  $J(Y;X) = \overline{p_Y^T} \, \overline{p_Y} = J(X;Y)$. Thus the scalar measure is symmetric.

- Now we check whether the scalar valued measure satisfies the triangular inequality ( The random variable X is the input to a discrete memoryless channel whose output is Y. Y is inturn the input to another discrete memoryless channel whose output is Z )

$$J(X;Y) = \overline{p_X^T} \, Q \, \overline{p_Y} = \overline{p_Y^T} \, \overline{p_Y},$$
$$J(Y;Z) = \overline{p_Y^T} \, Q \, \overline{p_Z} = \overline{p_Z^T} \, \overline{p_Z},$$

Hence we necessarily have that
$$J(X;Y) + J(Y;Z) = \overline{p_Y^T} \, \overline{p_Y} + \overline{p_Z^T} \, \overline{p_Z}.$$

But by definition $J(X;Z) = \overline{p_X^T} \, Q \, \overline{p_Z} = \overline{p_Z^T} \, \overline{p_Z}$.

Thus $J(X;Y) + J(Y;Z) \geq J(X;Z)$.

Hence the triangular inequality is satisfied. Thus, the following Lemma is established.

**Lemma 2:** The scalar valued measure
$$J(X;Y) = \overline{p_X^T} \, Q \, \overline{p_Y} = \overline{p_Y^T} \, \overline{p_Y} \ldots\ldots\ldots(2.15)$$

between the probability vectors ( corresponding to the probability mass functions of the random variables X, Y ) is a "Pseudo—Metric" on the space of probability vectors ( Where the random variable Y is output of a Discrete Memoryless Channel whose input is X ).

Now we summarize the results discussed so far in the following:

- $J(X) = [\sum_{j=1}^{\infty}[p_X(j)]^2]^{\frac{1}{2}}$ is like the "Euclidean Length" of a probability vector. In Lemma 1, it was shown that $1 - [J(X)]^2$ approximates the entropy of the random variable X.
- J(X;Y) is a scalar valued measure on the input X and output Y of a discrete memoryless channel.

**4. Conclusions**:

In this research paper, the relationship between Gibbs-Shannon Entropy measure and the Tsallis entropy ( for q=2 ) is demonstrated. Based on this result, various interesting measures associated with probability mass functions (defined at the input and output of a Discrete Memoryless Channel ) are defined. It is expected that these results will be of utility in Information Theoretic Investigations.

**REFERENCES:**

[Ash] R.B.Ash,"Information Theory," Dover Publications, Inc, New York,

[Rama] G. Rama Murthy, "Weakly Short Memory Stochastic Processes: Signal Processing Perspectives, " Proceedings of International Conference on Frontiers of Interface between Statistics and Sciences," December 20, 2009 to January 02, 2010,